\documentstyle[mprocl]{article}

\def\Journal#1#2#3#4{{#1} {\bf #2}, #3 (#4)}

\def\PLB{{\em Phys. Lett.}  B}

\def\PRD{{\em Phys. Rev.} D}

\def\be{\begin{equation}}
\def\ee{\end{equation}}
\def\bea{\begin{eqnarray}}
\def\eea{\end{eqnarray}}
\begin{document}

\title{MATHEMATICAL PROBLEMS IN HIGHER ORDER GRAVITY AND COSMOLOGY}

\author{S. COTSAKIS}

\address{Department of Mathematics\\University of the Aegean\\Karlovassi 83200, Samos
\\Greece}
\maketitle\abstracts{We discuss the issue of motivating the analysis of higher
order gravity theories and their cosmologies and introduce a rule which states
that these theories may be considered as a vehicle for testing whether certain
properties may be of relevance to quantum theory. We discuss the physicality
issue arising as a consequence of the conformal transformation theorem, the
question of formulating a consistent first order formalism of such theories and
also the isotropization problem for  a class of generalized cosmologies. We
point out that this field may have an important role to play in clarifying
issues arising also in general relativity.
}
\section{Motivation.}
We  consider theories of gravity of the general nonlinear lagrangian type
 \be
L_{g}=f(r),
\ee
where $r=R,\; Ric^2 ,\; Riem^2$,
called collectively {\em higher order gravity theories (HOG)}. Doing
cosmology with such actions leads to {\em 
higher derivative  cosmologies (HDCs).}

The field equations arising from these $f(r)$--actions are
typically of fourth order.
Motivating the study of such systems is non--trivial and indeed there have been
mixed feelings in the literature concerning this issue.
Among the often quoted virtues of adopting such actions include
 the fact that they
could constitute a
  first approximation to a non-existent quantum theory of
gravity, represent a kind of  unifying theory,
and also cure `problems' of GR/Cosmology such as providing
better singularity behaviour (singularity avoidance)
and/or   better late--time cosmological behaviour. In fact,
 there is
no rule to force the form of the gravitational action to be necessarily of
the usual Einstein--Hilbert type.
Adopting this philosophy, one typically ends up with a
larger  solution space than that of GR. This in turn raises several issues, for
instance that of  understanding the  precise
relation between the two solution spaces and also
 the  structure near the singular (conformal--see below) boundaries.

On the other hand, there are several remarks one could make
on the negative side of things.
 Here are some examples:
 Some authors
simply hold  that such actions lead to  unphysical choises for the
gravitational field and as such they
cannot be {\em seriously} considered as a viable approximation  to
quantum theory. Moreover, it may be thought that the whole issue of assuming
such a drastic alteration in the form of the gravitational action is not a
`burning' one in view of the fact that, for example, strictly speaking the
issue of singularities in GR in not a closed one and one should clarify this
and related matters first. Lastly, it should be pointed out that the
  initial value problem is not well--posed for some
theories.

         To proceed, we introduce the following alternative motive for
considering such theories which we call the
{\em Universal Admittance Rule:}
If a certain property is valid in GR {\em and} in
all other physically interesting theories then this may well be an
indispensible ingredient of a more fundamental theory (eg, Quantum Gravity).

We believe that under the above rule it becomes meaningful
 to consider this kind of
variants to GR and indeed one may view such alternative  theories and their
cosmologies as a more friendly, testing ground
for discovering which {\em properties} may prove to be truly fundamental.
Such `properties' may be
   black hole entropy,
 inflation,
  isotropization property,
recollapsing property, questions of
  stability in cosmology,
hamiltonian structures/principles, etc.
\section{Physicality Issue.}
We start with the
conformal equivalence theorem cf. \cite{ba-co88}. This states that any of
these higher order systems may be regarded as GR with additional fields in
conformal space. For example, taking the starting lagrangian to be an analytic
function of the scalar curvature, $f(R)$,
 and performing a conformal transformation one
obtains
\be
 f(R)+T_{m}
\stackrel{\tilde{g}=\Omega^{2}g}{\Longleftrightarrow}
\tilde{R}+T_{\phi}+\tilde{T}_{m}.
\ee
An immediate advantage of this result
is that we have cast the original system in the
form of a symmetric hyperbolic system which is easier to analyze.
However,  since there are now {\em two}
metrics on $M$ the question naturally arises
as to which is the physical metric among $g,\tilde{g}$.
The following result shows that in certain cases only one of these metrics
 may be the true one
\cite{cot}.

{\bf Theorem.}
$\tilde{g}$ is always the physical metric for certain manifolds 
provided $\nabla_a\nabla_b\phi =0$.

The proof consists in constructing the  types of admissible manifolds
by introducing  and exploiting the consequences of a   generalized form
of a theorem due to Bochner  (cf. \cite{cot}).
It also involves an analysis of the
behaviour of spacetime metrics with $Ric <0$ which is, in general, a
very delicate,
 subtle,  interesting and open (not completely settled even for Riemannian
manifolds--not to mention spacetimes)  question.

\section{Constrained Variations and Conformal Structure.}
Developing a first order formalism for higher order systems of the form
discussed above is not a closed issue and in fact these methods (and more
generally those of a metric--connection type) may provide us with an
 alternative to reduction of order.
We  \cite{cmq97} have recently shown that the field equations  obtained
from varying the
most general, pure--metric, higher order lagrangian with general matter
couplings and with an arbitrary symmetric connection $\nabla $,
$L\left( g,\nabla g,...,\nabla ^{\left( m\right) }g;\,\psi ,\nabla \psi
,...,\nabla ^{\left( p\right) }\psi \right)$ are equivalent to those found via
the Palatini variation of the metric--connection lagrangian
$L^{\prime }\left( g,\Gamma ,\Lambda ,\psi \right) =L\left( g,\Gamma ,\psi
\right) +L_c\left( \Lambda ,\Gamma \right)$.
A consequence of this result is the following
 generalization of the conformal equivalence theorem.
Consider the so--called   Weyl geometry
wherein $\nabla _cg_{ab}=-Q_cg_{ab}$
with $Q_c$ the Weyl covariant vectorfield. Then upon a conformal transformation
 the $f(R)$ field equations in Weyl geometry reduce to the form
\begin{equation}
\widetilde{G}_{ab}=\widetilde{M}_{ab}^Q-\widetilde{g}_{ab}V\left( \varphi
\right) ,  \label{WCeqn}
\end{equation}
where
\[
\widetilde{M}_{ab}^Q=-\widetilde{\nabla }_{\left( a\right. }\widetilde{Q}%
_{\left. b\right) }+\widetilde{Q}_a\widetilde{Q}_b+\widetilde{g}_{ab}\left( -%
\widetilde{Q}^2+\widetilde{\nabla }^m\widetilde{Q}_m\right) . 
\]

Notice that if the
 geometry is Riemannian, i.e. $\widetilde{Q}_a=0$ (original Weyl vector
 is a gradient, $Q_a=\nabla_a\Phi $) this generalised system is reduced to the
usual one.

\section{Isotropization Theorem for HDCs.}

Can the present isotropic state arise from `arbitrary' initial conditions?
A first  answer to this question is contained  in a well--known theorem of
 Collins and Hawking \cite{co-ha73}:
The  set of spatially homogeneous cosmologies that
can approach isotropy at late times is of measure zero in the space of all
spatially homogeneous initial data.

The corresponding $f(R)$--isotropization problem
may not be obtained  directly since  in the corresponding  Raychaudhuri equation
the Ricci term comes from the $f(R)$ field equations.
However, if we conformally transform the Raychaudhuri equation we obtain a
{\em Raychaudhuri system} (generalization of the usual Raychaudhuri equation)
and noting that in our case
the  potential
is   not necessarily globally convex, we arrive at an
isotropization result of
a  Collins--Hawking    type that is, that {\em their
 theorem I is valid in higher order gravity theories ie, Bianchi types I,
V and VII approach an isotropic state.}
\section{Future Work.}
The field discussed in the present paper certainly contains a host of
well--defined problems to consider and these  may  prove
to be fruitful avenues of research in the near furure on problems related to
the definition of the tendency of cosmological spacetimes to isotropize,
recollapse, the role of cosmic no--hair conjecture, their conformal hamiltonian
structure, their Cauchy problem and  the role of black hole entropy.

\end{document}